\documentclass[12pt,preprint]{aastex}

\begin{document}

\title{IR Detection of Low-Mass Secondaries in Spectroscopic Binaries}

\author{{Tsevi Mazeh\altaffilmark{1,2}, L. Prato\altaffilmark{1,3,4},
M. Simon\altaffilmark{1,3,5},}\\
Elad Goldberg\altaffilmark{2},
Dara Norman\altaffilmark{1,6},
and Shay Zucker\altaffilmark{2}}

\altaffiltext{1}{Visiting Astronomer at the Infrared
Telescope Facility, which is operated by the University of Hawaii under
contract to the National Aeronautics and Space Administration.}
\altaffiltext{2}{Department of Physics and Astronomy, Tel Aviv University,
Tel Aviv, Israel; mazeh@wise.tau.ac.il}
\altaffiltext{3}{Visiting Astronomer, Kitt Peak National Observatory,
National Optical Astronomy Observatories, operated by the Association of
Universities for Research in Astronomy, Inc., under cooperative agreement
with the National Science Foundation.}
\altaffiltext{4}{Department of Physics and Astronomy,
University of California, Los Angeles, CA, 90095-1562; lprato@astro.ucla.edu}
\altaffiltext{5}{Department of Physics and Astronomy,
SUNY$-$SB, Stony Brook, NY 11794-3800; michal.simon@sunysb.edu}
\altaffiltext{6}{Cerro Tololo Inter-American Obs., Casilla 603,
La Serena, Chile}

\begin{abstract}

This paper outlines an infrared spectroscopic technique to measure the
radial velocities of faint secondaries in known single-lined binaries.
The paper presents our $H$ band observations with the CSHELL and the
Phoenix spectrographs and describes detections of three low-mass
secondaries in main-sequence binaries: G147-36, G 164-67, and HD 144284,
with mass ratios of $0.562 \pm 0.011$,\ $ 0.423 \pm 0.042$, and $0.380
\pm 0.013$, respectively. The latter is one of the smallest mass
ratios derived to date.

\end{abstract}

\keywords{binaries: spectroscopic---infrared: stars---techniques: spectroscopic}

\section{Introduction}

This paper describes a technique to measure the radial velocities of
faint secondaries residing in known single-lined spectroscopic
binaries (SB1s) using high-resolution infrared (IR) spectroscopy.  The
paper outlines the approach and reports on the detection of low-mass
secondaries in three main-sequence binaries. From the radial
velocities of the faint secondaries, combined with the already known
velocities of the primaries, we derive the mass ratios of these
binaries, and estimate the masses of the secondaries.

In most of the known spectroscopic binaries only the primary spectrum
is detected and measured, rendering such systems as SB1s.
Measurements of SB1s yield only the mass function, in which the mass
of the primary, secondary, and inclination ($M_1$, $M_2$, and $i$) are
inseparable.  To derive the mass ratios of these systems it is
necessary to turn them into double-lined binaries (SB2s). To obtain
the masses we further need to obtain information about the
inclination, either from the light curves in eclipsing binaries, or
from astrometric orbits.  When the inclination of an SB2 is not known,
it is possible to derive the secondary mass if the primary mass can be
estimated from the system's photometry and spectrum.

Accurate mass ratios and masses are important for many aspects of the
astrophysics of binaries.  We are applying the technique we describe
here in two areas of particular interest to us.  Accurate masses for 
pre$-$main-sequence (PMS) stars
with low masses (M$<$1 M$_{\sun}$) are known for only very few
objects.  As a result, calculations of PMS evolution are incompletely
tested.  Comparison of evolutionary tracks published in the past 5
years show that, for a given location of a star in the HR diagram, the
different tracks can produce a scatter of 2$-$3 in the inferred
mass and age \citep{sim00}.  One of our goals is, therefore, to
improve the calibration of the theoretical tracks by deriving the mass
ratios, and, eventually, the masses of the low-mass components of PMS
binaries.

Another of our goals is to derive the distribution of mass ratios and
secondary masses in main-sequence binaries.  The secondary-mass
distribution touches upon one of the unanswered basic questions of
star formation, namely how binaries, and close binaries in particular,
are formed.  The secondary distribution provides one way to
distinguish \citep{bos93} between the star formation models
(e.g. Clark 1995; Bate 2000).  In addition, the lower end of the
secondary mass distribution is of crucial importance to an issue that
has attracted recent attention: the distinction between low-mass
secondaries and extrasolar planets \citep{han00, pou01a, pou01b,
zuc01a}. Some recent studies \citep{bas97, maz99, hal00, jor01, zuc01b}
suggested that the two populations could be separated by comparing
their mass distributions.

Spectroscopic binaries have been studied mostly in the visible, where
sensitive detectors have been available for a few
decades. Unfortunately, detection of secondaries in visual spectra
favors systems with light ratio, and hence mass ratio, close to unity.
Most of the known spectroscopic binaries are thus SB1s and the
distribution of mass ratios in the known double-lined systems (SB2s)
is biased toward unity.

Our approach is based on the premise that in each binary the secondary
has a lower mass and is therefore cooler than the primary, favoring
its detection in the IR.  Consider, for example, a binary with 0.2 and
1.0 M$_{\sun}$ stars at the age of 5 Gyr. The results of \citet{bar98}
and \citet{hen90} indicate that the flux ratio between the primary and
the secondary is about $8\times10^{-4}$ at 5500 \AA, but only 0.01 at
1.6~$\mu$m.  Thus, the advent of high-resolution IR spectrometers with
efficient large format detectors offers the means to complement the
visible radial-velocity studies with IR observations that are
considerably more sensitive to low-mass companions. This enables the
measurement of the secondary radial velocities which convert the SB1s
into SB2s.  The derivation of mass ratios by this approach does not
require extensive observations since the parameters of the primary
orbit have already been derived from previous work so only a
small number of secondary velocities are required.

To analyze the composite target spectra we use TODCOR, a two-dimensional
correlation algorithm developed by \citet{zuc94}.  The
algorithm assumes that the observed spectrum is a combination of two
known spectra, shifted by the radial velocities of the
two components, and calculates the correlation of the observed
spectrum against a combination of two templates with different shifts.
TODCOR generates a two-dimensional correlation function, whose
peak simultaneously identifies the radial velocities of both the
primary and the secondary (e.g. Zucker \& Mazeh 1994).  One of the
advantages of TODCOR is its ability to use different templates for the
primary and the secondary. When the primary and the secondary are of
different spectral types, the use of two different templates enables
us to utilize all the spectral information contained in the
observed spectrum of the combined system.  This feature of TODCOR is
most important when deriving the radial velocity of a faint secondary.

This paper outlines the new approach and reports on the detection of
low-mass secondaries in three main-sequence binaries: G~147-36,
G~164-67, and HD~144284.  The former two systems were identified and
studied as SB1s by \citet{lat01}, who searched for spectroscopic
binaries within a sample of high-proper-motion stars \citep{car87}.
The latter was found by \citet{duq91}, who studied the nearby G stars.
The three stars presented here were the three most significant
detections in our main-sequence program to date.  Preliminary reports on the
application of this technique to main-sequence and PMS SBs appear in
\citet{maz00}, \citet{pra01a}, \citet{ste01}, and \citet{pra01b}.
Section 2 presents four approaches to the derivation of mass ratios
from the primary and secondary radial velocities.  Our observations
are described in Section 3 and the spectral type templates in
Section~4.  We present our derivation of the radial velocities in
Section 5, while Section~6 shows the calculation of the mass ratios
for the three binaries. A short discussion of our results appears in
Section~7.

\section{The Derivation of the Mass Ratios}

In this section we present four different approaches to derive the
mass ratio $q=M_2/M_1$ of a spectroscopic binary from a few radial
velocities of both the primary and the secondary. Choosing between
these approaches depends upon the extent of the information available
from the previous SB1 solution.

\begin{enumerate}

\item When no previous orbital information is available, any two
{\it pairs} of velocities can yield the mass ratio of a binary (Wilson
1941). Suppose we have observed the system at two epochs, $t_i$ and
$t_j$, with velocities $\{V_{1,k}, V_{2,k}; k=i,j\}$ where $V_{1,k},
V_{2,k}$ are the velocities of the primary and the secondary,
respectively. Then the mass ratio is

 $$   q_{i,j} = -{{V_{1,i} - V_{1,j}}\over{V_{2,i} - V_{2,j}}}\ . $$

The ratio of the velocity {\it difference}
between the i-th and the j-th observations is exactly $q$.
This approach will be most effective with measurements made at such
phases that the difference of the velocities is large compared to the
uncertainty of their measurements.  When more than two pairs of
velocities are available, any two pairs of velocities can yield
another value of the mass ratio. Note, however, that not {\it all} the
$q$ values are independent (see also Wilson 1941).

\item When the systemic velocity, $\gamma$, is available
from the SB1 observations, each pair of velocities, obtained at time
$t_i$, can yield a value of the mass ratio

$$q_{i} = - {{V_{1,i} - \gamma} \over {V_{2,i} - \gamma}}\ .  $$

\item When the elements of the single-lined orbit are available, the
expected primary velocity $V^{exp}_{1,i}$ at the time of the
observation $t_i$ can be derived from the elements. In such a case,
each secondary velocity $V_{2,i}$ can yield a value of the mass ratio

$$q_{i} = - {{V^{exp}_{1,i} - \gamma} \over {V_{2,i} - \gamma}}\ .  $$

In the case in which the zero point of the new velocities is uncertain, {\it
and} the primary velocity was also measured, $V_{2,i}$ can be replaced
by $V_{2,i}+(V^{exp}_{1,i} - V_{1,i})$.
This assumes that
the new primary and secondary velocities share the same zero point.

\item In the case in which the original single-lined radial velocity
measurements are available, one can combine the old velocities with
the new double-lined measurements and obtain a new solution for the
orbital elements. The new set of elements includes $K_2$, the
semi-amplitude of the secondary, from which one can derive the mass
ratio.

\end{enumerate}

The three binaries considered here have known elements
together with the individual measurements \citep{duq91, lat01}. We
therefore opted to apply the second and the last approaches listed
above. The advantage of the second approach is that the spread of the
mass ratio values can be examined and evaluated directly, while in the
last option this is all hidden in the orbital solution.

\section{Observations}

To search for the spectroscopic signature of the secondaries in our
SB1 sample, we observed the target systems and standard stars in 1997
December and 1999 February, May, and November with CSHELL on the IRTF
3-m and in 1999 December with Phoenix at the KPNO 4-m.
CSHELL is a high resolution IR spectrometer equipped with a
$256\times256$ InSb array that provided a free spectral range of
$c\Delta\lambda/\lambda \sim 730$ km/s.  We observed with a $0.5''$
slit, yielding a measured resolution of $\sim 30,000$.  Phoenix,
NOAO's high resolution IR spectrometer, has a $512\times 1024$ InSb
array which provides a $\sim 1440$ km/s free spectral range.  With
Phoenix, we used a $0.8''$ slit which gave a measured resolution of
$\sim 35,000$.

The spectrometers were centered at $1.5548~\mu$m for all of our
observations.  Inspection of atlases of solar photospheric and sunspot
spectra showed that this region contains lines suitable for the
characterization of stars in the mass range of interest. The region
includes atomic lines for identifying the G and early K spectral type
stars and molecular lines, most notably OH and H$_2$O, suitable for
late K and M spectral type stars \citep[see also Meyer et al. 1998 for
an overview of stellar spectra in the $H$ band]{liv91, wal92}.  This
wavelength region is also advantageously free of strong terrestrial
absorption lines.

We calculated the dispersion and wavelength calibration of
CSHELL and Phoenix at the $1.5548~\mu$m setting using spectral lamps
internal to the instruments.  Both spectrometers showed small flexure,
at the level of $\sim 3$ km/s during a night of observations.  We
corrected for the flexure by reobserving the lamps during the night and
also by monitoring the apparent positions of the OH lines detected from
the night sky.  The spectra were extracted by procedures written by one
of us \citep{pra98}.

\section{Templates}

In addition to the target spectra, we also observed a sample of
main-sequence standard stars, from spectral type F8 through M6.5, in
order to provide templates for the two-dimensional cross-correlation
analysis using TODCOR. When possible, we chose a standard whose
metallicity would be close to solar, on the expectation that its lines
would be deeper than those of older stars. The templates are usually
slow rotators, and if needed we generated new calculated templates by
broadening the template lines with profiles representing different
rotational velocities (see below).  We also expected that these
standards would provide better templates for our PMS star program.

Table~1 lists the observed standards.  The third column provides an
estimate of their population type.  A numerical entry in this column
represents the value of [Fe/H] from the catalog of \citet{cay92}.  In
cases where the catalog listed several determinations, we used the
most recent one.  A dagger ($\dag$) in this column is \citet{kee89}'s
designator for stars they judged to have the best determination of
spectral type.  OD and YD are Leggett's (1992) ``old disk'' and
``young disk'' designations, based on kinematic and photometric
criteria.  Column 4 lists the assigned radial velocities of the
observed spectra (see below). Column 5 indicates the instrument used,
CSHELL (C) or Phoenix (P), on the date indicated in Column 6.

Figure~1 shows all the template spectra measured with the Phoenix
spectrometer.  The spectra are shifted to the laboratory frame.  Some
of the prominent lines are indicated following the identifications
given by \citet{liv91} and \citet{wal92}.  Since the TODCOR analysis
correlates spectra of the templates and targets, only the presence of
lines matters; their specific identification is unimportant.  The
sawtooth pattern evident in the spectrum of HR~1543 is an artifact of
the illumination of the array for that particular observation.  As
Table~1 indicates, we observed a similar range of spectral types, and
often the same stars, with CSHELL and Phoenix.  In Figure 2, we show
the CSHELL and Phoenix observations for the M stars only.  Comparison
shows that most of the features in the complex spectra of the late M
types, which appear to be noise, are in fact real.

The radial velocities of the observed spectra of the standards are
crucial for our analysis.  As a starting point, we assigned to each
spectrum the radial velocities of these template stars measured either
by the Geneva \citep{duq91} or by the CfA groups \citep{lat85}.
However, not all our templates were observed by these two
groups. Moreover, some of the observed spectra of the templates were
shifted by about 1--2 km/s, apparently the result of random jitter of
the spectrographs. To overcome these two difficulties we performed a
few iterations to rederive new assigned velocities for each of the
observed template spectra. This was done by correlating each observed
spectrum with templates of similar classes, using the assigned
velocities of the previous iteration. A few iterations led us to
assign a velocity for each observed spectrum as given in Table~1.

The assigned velocities are {\it not} meant to measure the actual
radial velocities of these stars. They are our best estimates for the
radial velocity shifts of the observed spectra, and therefore these
are the values we used in our analysis with TODCOR.  In an ideal
world, without any outliers and spectrograph jitters, these velocities
would be our best estimate for the actual stellar radial
velocities. Actually, we find that for most of the stars for which we
had a prior radial velocity measurement \citep{duq91, lat01}, our
assigned radial velocity was within 1$-$2 km/s of the previous
measurement. We therefore estimate that our zero-point is good to
within 1.5 km/s.  Only one star, GJ 475, displayed a discrepancy of 6
km/s between the CSHELL (8.0 km/s) and Phoenix (1.9 km/s)
measurements. \citet{duq91} derived a velocity of $6.39 \pm 0.08$
km/s. We, nevertheless, used the two spectra with the assigned
velocities, assuming some non-random mechanical or optical shift of
4.5 km/s of the Phoenix instrument had occurred during that exposure.

\begin{table}[t]
  \caption[]{Template Library}
  \label{tab:par}
  \begin{center}
\begin{tabular}{|l|lrrcr|}
\hline
\hline
Star   &Sp Ty& Pop& V$_{rad}$& Instr.& Date(JD)\\
       &     & Type   &  km/s   &       & 2450000+\\
\hline
HR 1543& F6  & +0.02&       24.5 &   C   &  0806.93\\
       &     &      &       20.5 &   P   &  1532.88   \\
HR 1780& F8  &      &       38.1 &   C   &  0806.95   \\
       &     &      &       36.1 &   C   &  1228.75   \\
       &     &      &       35.0 &   P   &  1530.61   \\
GJ 475 & G0  & -0.19  &      8.0 &   C  &   0806.06   \\
       &     &      &        1.9 &   P   &  1533.06 \\
HR 88  & G2.5& +0.20  &     -5.4 &   P   &  1532.63  \\
HR 2643& G4  & -0.15  &     24.5 &   C   &  0806.97  \\
HR 5072& G4  &$\dag$  &      3.2 &   C   &  1300.92 \\
HD 37986&G8  &      &       59.0 &   C   &  1228.79   \\
HR 3259& K0  &       &      29.7 &   C   &  0805.96   \\
GJ 315 & K1  &      &      -10.9 &   C   &  0804.07  \\
HR 493 & K1  & -0.20&      -35.6 &   P   &  1530.59   \\
GJ 483 & K3  &      &        9.0 &   C   &  1229.07   \\
HR 753 & K3  &$\dag$&       22.7 &   P   &  1530.61  \\
GJ 570A& K4  &+0.01 &       26.0 &   C   &  1229.02  \\
HR 8085& K5  &-0.06 &      -64.0 &   C   &  0804.68 \\
       &     &      &      -67.1 &   P   &  1530.58 \\
HR 8086& K7  & 0.00 &      -64.0 &   C   &  0804.68 \\
       &     &      &      -65.3 &   P   &  1529.62 \\
GJ 270 & M0  & OD   &      -68.9 &   C   &  0804.01 \\
GJ 382 & M1.5& YD   &       12.0 &   C   &  0806.00 \\
       &     &      &       12.0 &   C   &  1228.94  \\
GJ 447 & M4  & OD   &       -31.1&   C   &  0806.03 \\
       &     &      &       -31.1&   C   &  1228.07  \\
       &     &      &       -32.9&   P   &  1531.04 \\
LHS 292& M6.5& OD   &        0.0 &   C   &  0805.06 \\
       &     &      &        0.0 &   C   &  1300.81 \\
       &     &      &        1.8 &   P   &  1530.96 \\
\hline
\end{tabular}

\end{center}
\end{table}

\section{The Radial Velocities}

To derive the radial velocities of the two stars in each binary, TODCOR
requires two specific template spectra, one for the primary and the
other for the secondary.  We ran TODCOR for a variety of different pairs
of templates from our accumulated library to optimize the match between
the templates and the observed composite spectra.  To each pair
of templates we assigned, as a measure of its fit to the binary spectra,
the correlation peak obtained by TODCOR for that pair of templates,
averaged over all the observed target spectra.  We then chose the pair of
templates that gave the highest average value for the peak.

To improve the fit between the observed binary spectrum and the two
templates, we matched the rotational broadening of the primary and
secondary lines with those of the templates. This was done by
broadening the template lines with profiles representing different
rotational velocities, forming an extended template library.  We then
chose the broadened templates that best match the observed spectrum of
the binary. To extract the radial velocities for the two stars, the
brightness ratio of the two templates, $\alpha$, was also required.
Assuming that each observation of a given binary system had the same
brightness ratio, we determined a value to be used for all of its
spectra. To choose the best brightness ratio, we ran TODCOR for a grid
of selected $\alpha$ values, searching again for that value which gave
the highest averaged correlation peak. We do not consider
the templates and $\alpha$ values chosen by this procedure as
representative of the true spectral types and light ratios of the binary
targets because the template and target stars may differ in abundances
and, for PMS stars, surface gravity.  We regard the template spectral
types and $\alpha$ as a set of parameters that yields the best radial
velocities. Choosing the best templates and $\alpha$ comprised the 
major part of the first step of our procedure.  Only after the best 
templates and the
brightness ratio were determined did we proceed to derive the radial
velocities, given in Table~2.  Since the essential inputs to TODCOR
are the existence and locations of lines in the spectra of the
templates and targets, the fundamental outputs are the velocities of
the primary and secondary, and hence their mass ratio.  The templates
and light ratios used are necessary to derive the velocities but are
not intended as models of the target binary.

While all observed spectra yielded reliable primary velocities, not
all spectra gave reliable {\it secondary} velocities. This is because
the secondary peak was not always sufficiently prominent, depending on
several factors including the SNR of the observed spectrum, the
velocity difference between the primary and the secondary, the small
differences between the actual spectra of the primary and the
secondary on one hand and their templates on the other, and the light
ratios between the primary and the secondary. The first two can change
between one exposure and another, and can explain why we could derive
a reliable secondary velocity from one spectrum and not from the
other. For our analysis we chose only spectra that yielded cross
correlation functions for which we judged the secondary peak to be high
enough above the other spurious peaks of the correlation.  Our
somewhat stringent criterion led us to accept only 10 secondary
velocities for the three stars, all listed in Table~2.  Figures
3$-$5 depicts the correlation functions derived by TODCOR for G147-36,
G164-47, and HD 144842.  The velocity uncertainty for the former two
objects is 1.5 km/s and for the latter is 2 km/s, because of its high
rotational velocity.

\begin{table}[t]
  \caption[]{Direct Mass-Ratio Determination}
  \label{tab:par}
  \begin{center}
  \begin{tabular}{rcrrrrl}
    \hline \hline
  JD           & Inst. & $V_1$ && $V_2$ &&$q_i$ \\
 (2450000+)    &       &\multicolumn{4}{c}{(km/s)}  &   \\
\hline
\multicolumn{7}{c}{G147-36~~  $\gamma=46.00\pm0.13$\, km/s}\\
   807.06      & C     &  15.56  & &  104.83 &  & $0.52\pm0.03$\\
  1228.01      & C     &   5.52  & &  112.54 &  & $0.61\pm0.03$\\
  1529.96      & P     &  12.85  & &  107.84 &  & $0.54\pm0.03$\\
  1532.98      & P     &  63.51  & &    8.22 &  & $0.46\pm0.04$\\
\hline
               &       &         & &         &$\bar{q}=$&$0.54\pm0.02$\\
\hline
\multicolumn{7}{c}{G164-67~~ $\gamma=-3.47\pm0.22$\, km/s}\\
  1302.84      & C     &  8.49   && -30.57 &   & $0.44\pm 0.06$\\
  1529.90      & P     &  8.77   && -33.82 &   & $0.40\pm 0.05$\\
\hline
               &       &         &&      &$\bar{q}$=&$0.42 \pm 0.04$\\
\hline
\multicolumn{7}{c}{HD 144284 $\gamma=-7.74\pm.20$\, km/s }   \\
   806.15      & C     & -28.93  &&  45.48 &   & $0.40\pm0.04$\\
   807.15      & C     &   -8.48 &&   0.81 &   & $0.09\pm0.23$\\
  1228.13      & C     &    0.73 && -30.08 &   & $0.38\pm0.10$\\
  1484.71      & C     &  -27.09 &&  43.90 &   & $0.37\pm0.04$\\
\hline
               &       &         &&        &$\bar{q}$=&$0.38\pm0.03  $ \\
\hline
\hline

\end{tabular}
\end{center}
\end{table}

\section{The Mass Ratios}

\subsection{Derivation from Systemic Velocities}

Table~2 shows the results of deriving the mass ratios using only
the $\gamma$ velocities, drawn from the previous single-lined orbital
solutions. The $\gamma$ velocity appears as the first entry
for each binary.  The Julian date, instrument used,
component velocities, and derived mass ratio
are then listed for each individual
observation.  The last entry in each section is
the average value and uncertainty of the individual mass ratios.

\subsection{Global Solution for Binary Parameters}

In this subsection we present the full double-lined orbital solutions
for the three orbits, derived with our ORB code \citep{maz93,
maz01}. To derive the double-lined solution, we assign different
uncertainties to the primary and secondary velocities of the different
measurements.  We estimated that the uncertainty of the previous
velocities of G147-36 and G164-67 \citep{lat01} are 0.5 km/s, and
those of HD 144284 \citep{duq91} are 1 km/s. The latter somewhat
bigger uncertainty is attributable to the large rotational width of
the stellar lines. The scatter we find for these velocities (see
Table~3) is very close to these estimates.

The uncertainty of the secondary velocities depends strongly on the
free spectral range, the number and depth of the absorption lines,
and, mainly, on the light ratio between the primary and the
secondary. Using two different sets of data, obtained with CSHELL and
Phoenix, complicates the analysis. Therefore, our uncertainty
estimates are poorly constrained.  We estimated that the uncertainty
of our secondary velocities for G147-36 and G164-67 are 1.5 km/s and
those of HD 144284 are 2 km/s, again because of the large rotational
velocity of the latter.

Table~3 presents our results, giving the derived values of the period,
P, in days, eccentricity, e, longitude of the periastron, $\omega$, in
degrees, time of periastron passage T$_\circ$, in J.D.$-$2440000,
primary velocity amplitude, $K_1$, in km/s, and the center-of-mass
velocity, $\gamma$, in km/s.  The only new parameter coming from the
secondary radial velocities is the velocity amplitude of the
secondary, $K_2$, in km/s, from which we derived the mass ratio $q$.
Finally, Table 3 lists $\sigma_{1,old}$, $\sigma_{1,new}$ and
$\sigma_{2,new}$, in km/s --- the r.m.s.\ of the residuals of the old
and new velocities of the primary, and the new secondary velocities.
The orbital elements derived here are entirely consistent with the
original published values.  The orbits are plotted in Figures 6$-$8.

The scatter for the primary and the secondary velocities, listed in
Table~3, gives an external estimate of our precision. As
can be seen from Table~3, these estimates are based on extremely small
numbers, only two measurements in the case of G164-67. It is therefore not
surprising that the secondary radial-velocity scatter ranges from
4.0 to 0.23 km/s. Our estimates for the uncertainty of the secondary
velocities are between these two values.

Table~4 is a summary of the mass ratios and masses estimated for the
three binaries. The primary masses for G147-36 and G164-67 are from
\citet{car94} and for HD 144284 from \citet{maz97}.
The latter is based only on the spectral type of the
system. Arbitrarily, we assigned an uncertainty of 0.1 $M_{\sun}$ to these
primary mass estimates.  The table also lists the minimum mass ratios
derived from the SB1 orbits. Statistically, the derived mass
ratios are expected to be close to the minimum values, but never
smaller. This is indeed the case, within the $1\sigma$ range, for all
three binaries.

\begin{table}[t]
  \caption[]{Orbital Elements for the three new SB2}
  \label{tab:par}
  \begin{center}
  \begin{tabular}{lrrr}
    \hline \hline
                & G 147-36    & G 164-67   & HD 144284 \\

\hline

P       &     6.572037   &   16.19429   &   3.0708216  \\
\medskip
$\pm$   &     0.000027   &    0.00067   &   0.0000069  \\

e       &     0.0051     &    0.2947    &   0.039      \\
\medskip
$\pm$   &     0.0030     &    0.0079    &   0.012       \\

$\omega$&        100     &      71.9    &      63      \\
\medskip
$\pm$   &         30     &       1.7    &      15       \\

T       &    7231.21     &    5785.327    & 5971.98      \\
\medskip
$\pm$   &       0.56     &     0.071    &    0.13       \\

$\gamma$&     46.329     &     -3.76    &   -8.23      \\
\medskip
$\pm$   &      0.090     &      0.11    &    0.20       \\

K1      &      44.04     &     19.53    &   25.10      \\
\medskip
$\pm$   &       0.13     &      0.16    &    0.31       \\

K2      &       78.4     &      46.2    &    66.0      \\
\medskip
$\pm$   &        1.6     &       4.6    &     2.2       \\

$\sigma_{1,old}$& 0.61     &      0.54    &     1.0      \\
\medskip
N   &         31     &        23    &      25       \\

$\sigma_{1,new}$&   2.8     &      0.57    &     1.7      \\
\medskip
N    &              4      &         2    &       4\\

$\sigma_{2,new}$&   4.0     &  0.23  &       2.7   \\
N           &        4     &     2  &         4   \\

\hline
    $\emph{q}$ & 0.562     & 0.423      &   0.380      \\
$\pm$          & 0.011     & 0.042      &   0.013      \\
    \hline \hline
  \end{tabular}
  \end{center}
\end{table}

\begin{table}[t]
  \caption[]{Masses of the Three New SB2's}
  \label{tab:par}
  \begin{center}
  \begin{tabular}{ccccc}
    \hline \hline
System  & $M_1$ &$q_{min}$ & $q_{derived}$& $M_2$  \\
\hline
G147-36  & 0.73 &   0.59  & 0.562  &  0.41 \\
\medskip
$\pm$    & 0.1  &   0.03  & 0.011  &  0.06  \\
G164-67 & 0.93  &   0.26  & 0.423  & 0.39   \\
\medskip
$\pm$   & 0.1   &   0.01  & 0.042  & 0.06   \\
HD 144284& 1.2  &   0.18  & 0.380  & 0.46   \\
\medskip
$\pm$    &  0.1 &   0.01  & 0.013  & 0.04   \\
\\
\hline

\end{tabular}
\end{center}
\end{table}

\section{Discussion}

The work presented here demonstrates the potential of IR spectroscopy
for detecting faint low-mass secondaries in single-lined spectroscopic
binaries and deriving their mass ratios.  We have shown that with the
CSHELL and the Phoenix spectrographs we can reach a mass ratio of
$0.38\pm 0.01$. To date, this is among the smallest known mass ratios
for a main-sequence spectroscopic binary. \citet{maz97}, observing in
the red part of the visual spectrum, derived mass ratios of
$0.57\pm0.02$ and $0.48\pm0.03$ for two binaries.

The detection of a secondary spectrum and the derivation of its radial
velocity is hindered by the presence of the bright primary in the
combined spectrum. For our purposes, the bright spectrum of the
primary is a strong background source that has to be removed from the
combined spectrum before we can measure the secondary radial
velocity. The subtraction of the primary leaves behind noise, which
ideally would be determined by Poisson statistics, but in reality is
determined by the imperfect match between the primary spectrum and its
template.  To reach the faint stellar secondaries, we have to exploit
every possible advantage available to improve the SNR of the secondary
spectrum.

In this work we have used two such advantages. We observed the binaries
in the IR regime, where the light ratio of the primary/secondary is
much smaller than in the visual, and we used the two-dimensional
cross-correlation analysis of TODCOR to extract the maximum amount of
spectral information from the observed spectrum.
To improve on this
further, we need high sensitivity IR spectra covering as large a
spectral range as possible.  Fortunately, IR echelle spectrometers
providing several orders simultaneously are now coming into operation
at the large aperture telescopes.  We have begun systematic
surveys to detect low-mass secondaries in both main-sequence and
PMS binaries \citep{pra01b}.

\acknowledgments

We thank W. Vacca (IRTF), R. Joyce (NOAO) and K. Hinkle (NOAO) for
thorough support of CSHELL and Phoenix and for advice and help in
their use. We thank J. Carr for advice about spectral type standards
and for sharing with us his list of standards.  We thank the referee
for an exceptionally informative and thorough report.  Our research made
extensive use of the SIMBAD database, operated at CDS, Strasbourg,
France. This research was supported in part by US-Israel Binational
Science Foundation Grant No. 97-00460 and by the Israel Science
Foundation, grant no. 40/00 (to T.M.), and by the US NSF, Grant
98-19694 (to M.S.).

\clearpage

\begin{figure}
\figurenum{1}
\epsscale{0.80}
\plotone{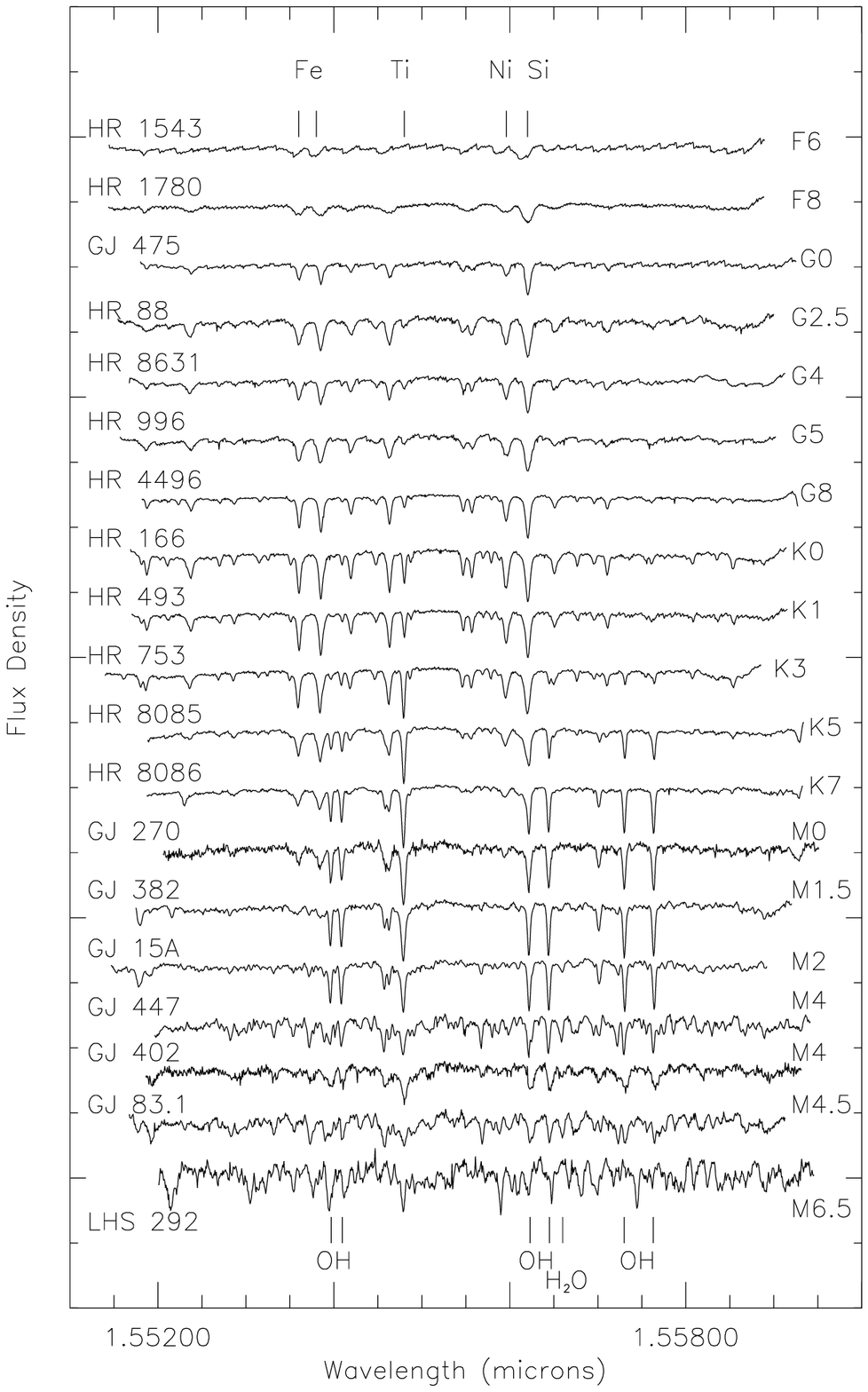}
\caption{Template spectra measured with the Phoenix spectrometer
at NOAO.  The spectra are shifted to the laboratory frame and flattened
by dividing by a second-order polynomial fit to the continuum.
The continuum level is normalized to 1; two divisions of the ordinate
correspond to the range 0 to 1.  The spectra are offset vertically
for clarity and a  few prominent lines are identified.}
\end{figure}

\begin{figure}
\figurenum{2}
\epsscale{0.80}
\plotone{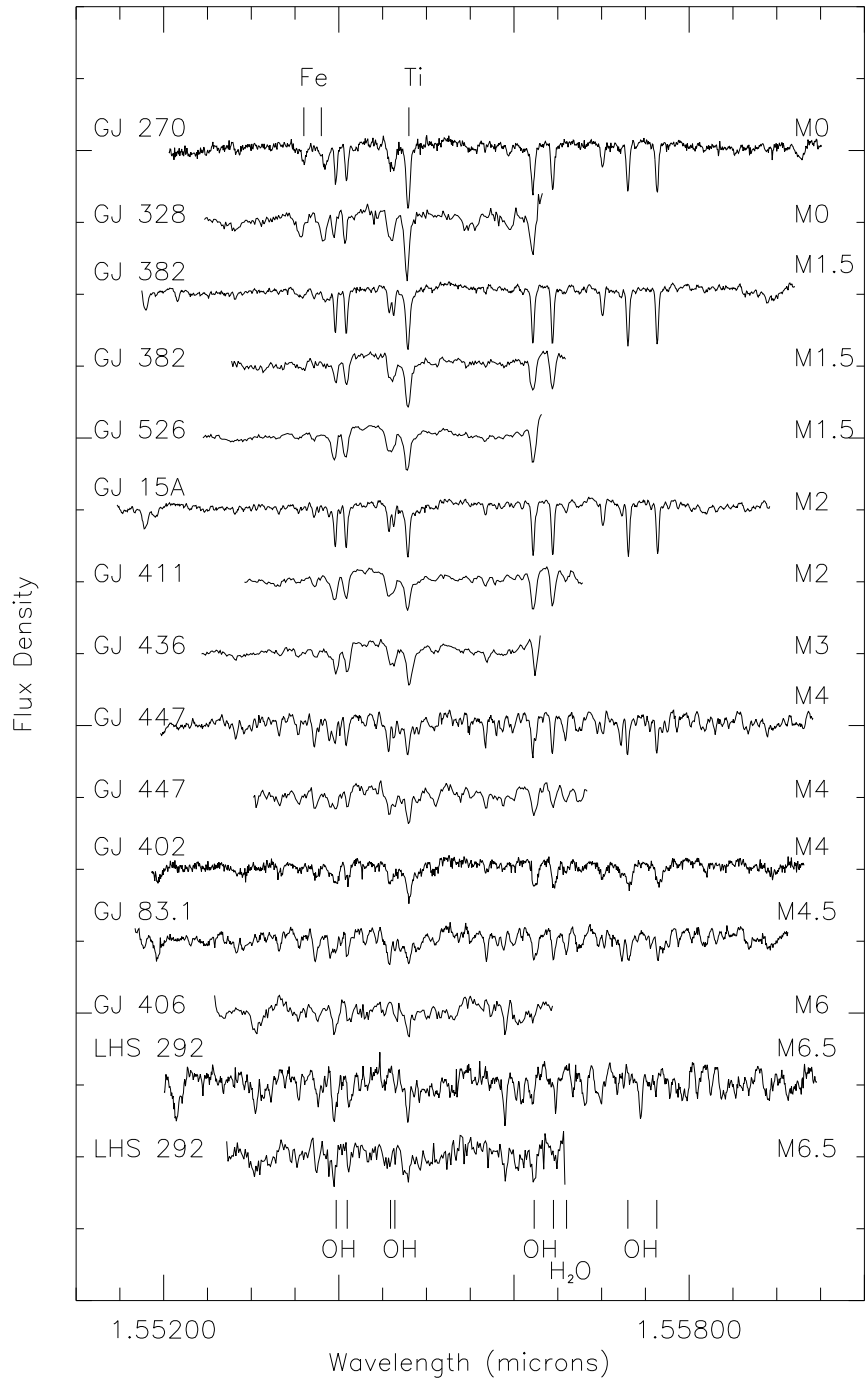}
\caption{Spectra of M stars obtained with Phoenix, as in
Fig. 1, and CSHELL, with a smaller spectral range, at the IRTF.  The
spectra are shifted in wavelength, flattened, normalized, and offset
as in Fig. 1.  Comparison of the late M type stellar spectra shows that 
most of the complex features, which
might be mistaken as noise, are in fact real.}
\end{figure}

\begin{figure}
\figurenum{3}
\epsscale{0.90}
\plotone{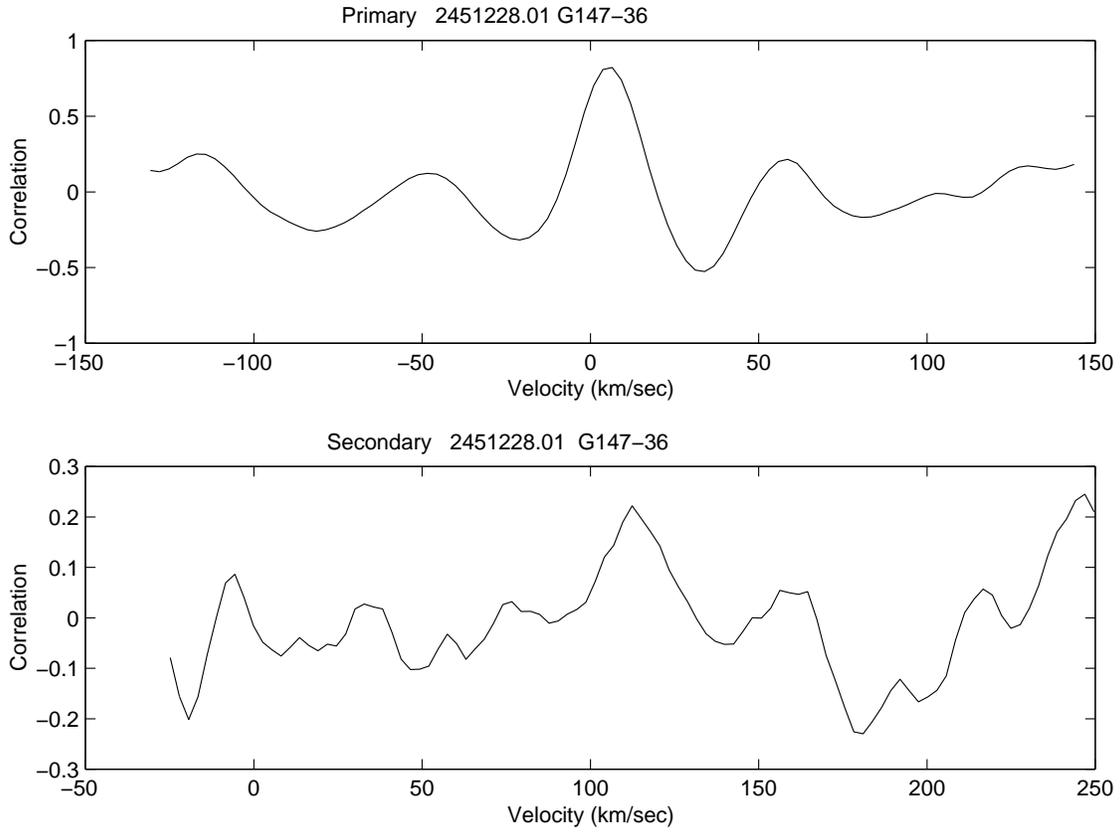}
\caption{The results of the TODCOR analysis for G147-36. The upper
panel shows the correlation of the observed spectrum with a template
for the primary. The lower panel shows the result of correlating the
template of the secondary with the difference between the observed
spectrum and the template of the primary, shifted to the derived
velocity of the primary.  The legends at the top of the panels give
the Julian date of the observations and the target name.}
\end{figure}

\begin{figure}
\figurenum{4}
\epsscale{0.90}
\plotone{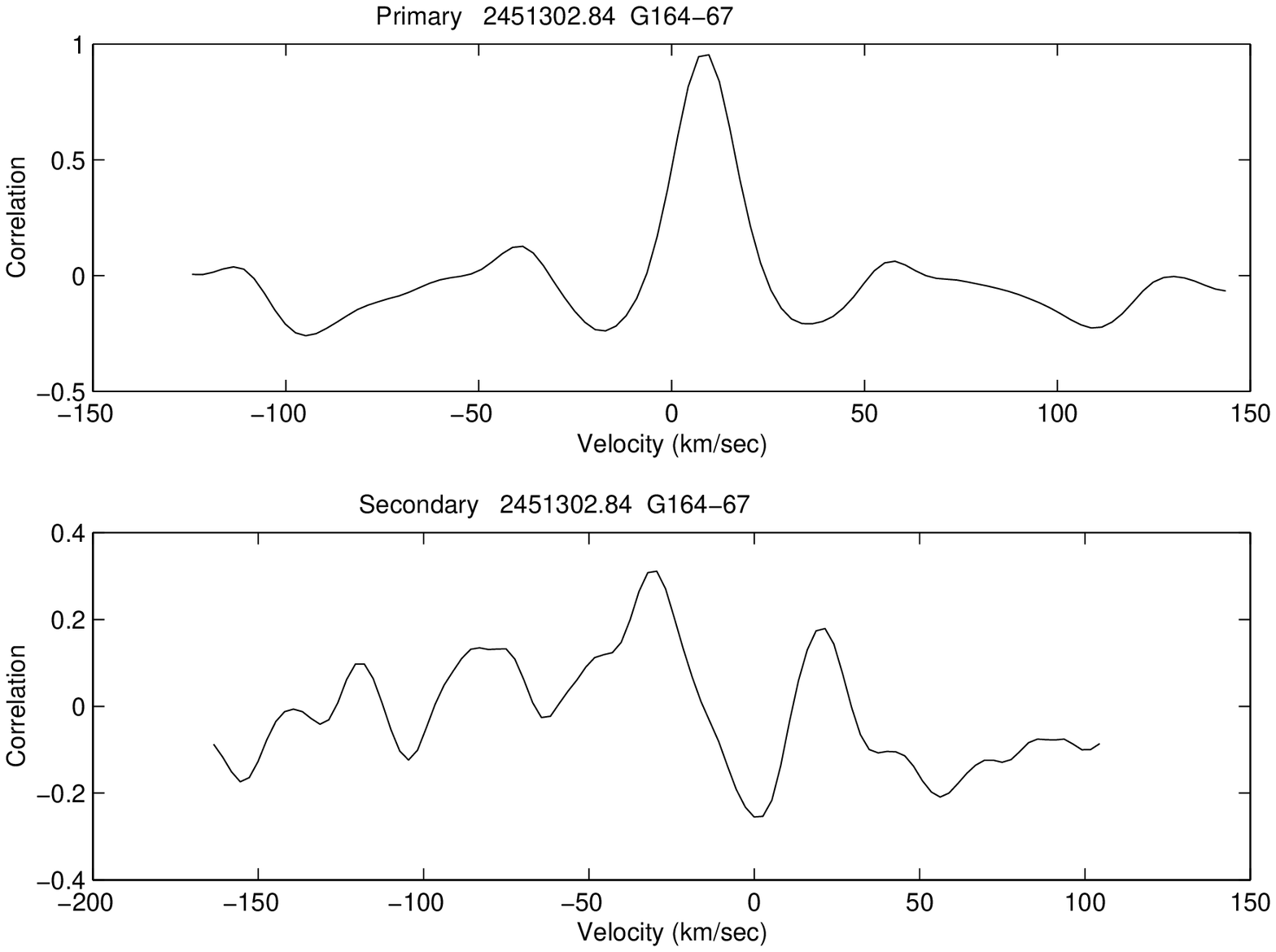}
\caption{Same as Figure 3 but for G164-67.}
\end{figure}

\begin{figure}
\figurenum{5}
\epsscale{0.90}
\plotone{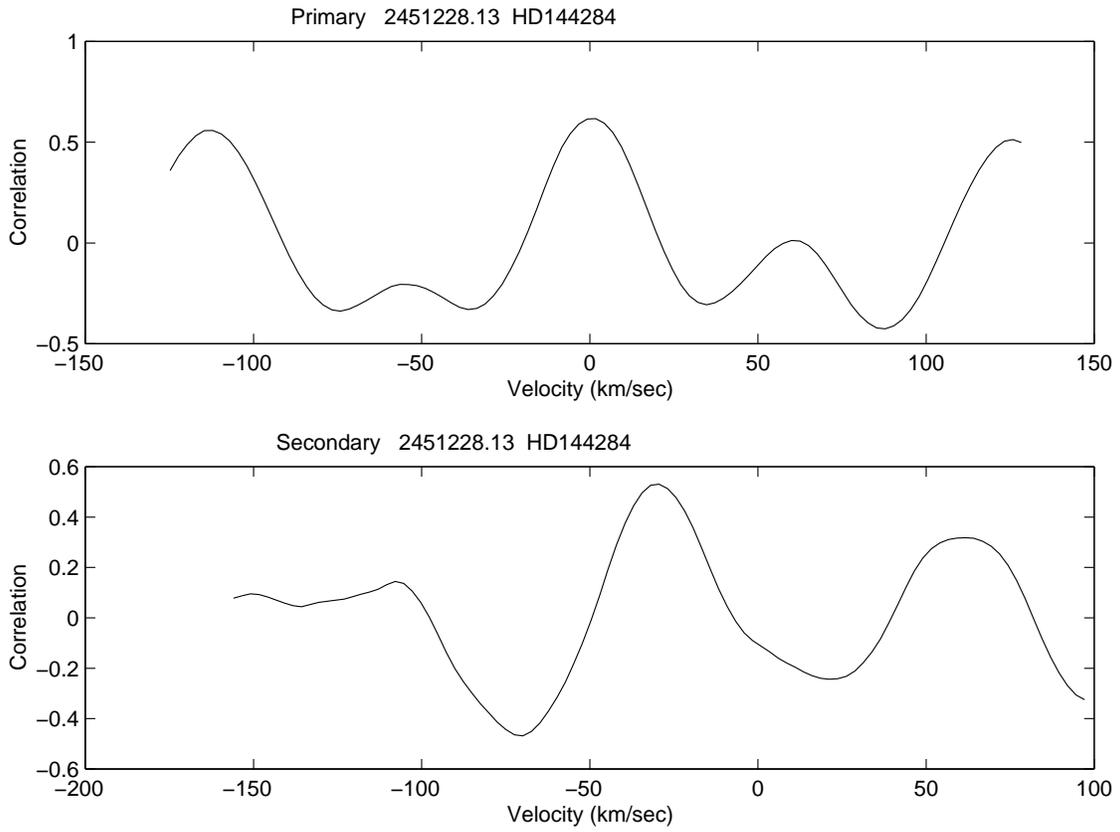}
\caption{Same as Figure 3 but for HD 144284.}
 \end{figure}

\begin{figure}
\figurenum{6}
\epsscale{0.90}
\plotone{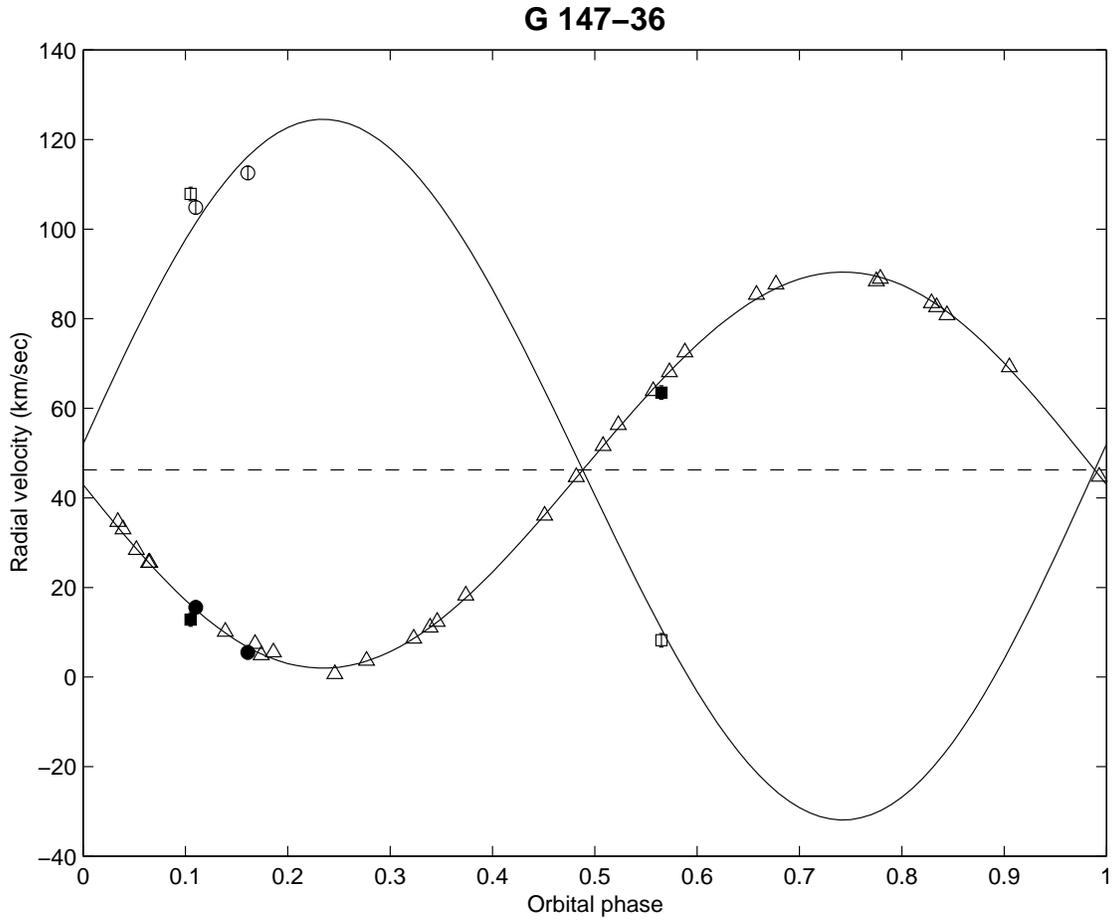}
\caption{The solid lines present the double-lined
spectroscopic orbital solution for G147-36.  The open triangles are
the previously measured velocities of the primary
\citep{lat01}. Measurements obtained with CSHELL are indicated by open
circles for the primary and closed circles for the secondary.  The
Phoenix measurements of the primary and secondary velocities are
indicated by the open and filled squares, respectively.}
\end{figure}

\begin{figure}
\figurenum{7}
\epsscale{0.90}
\plotone{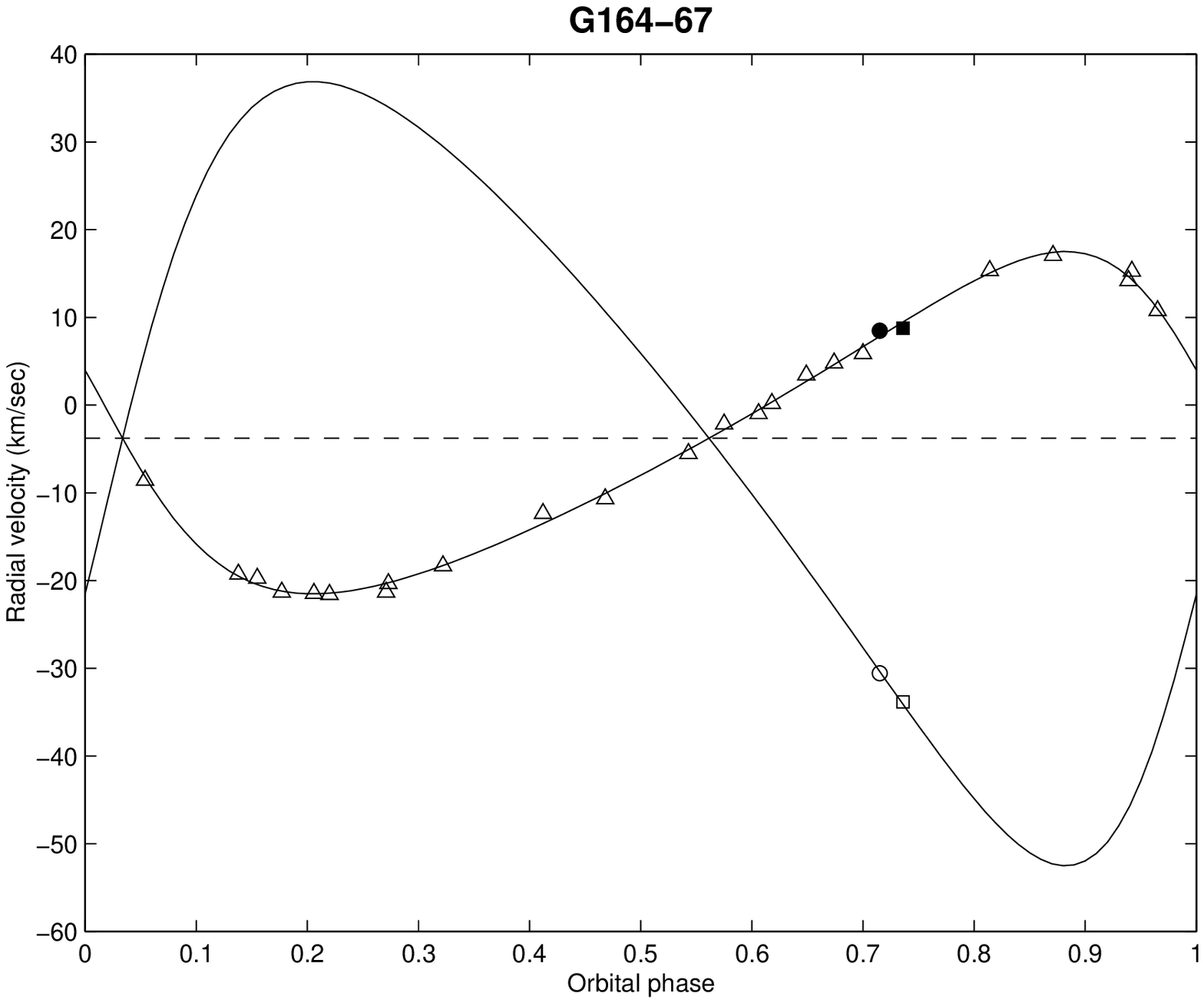}
\figcaption{Same as Figure 5 but for G164-67.}
 \end{figure}

\begin{figure}
\figurenum{8}
\epsscale{0.90}
\plotone{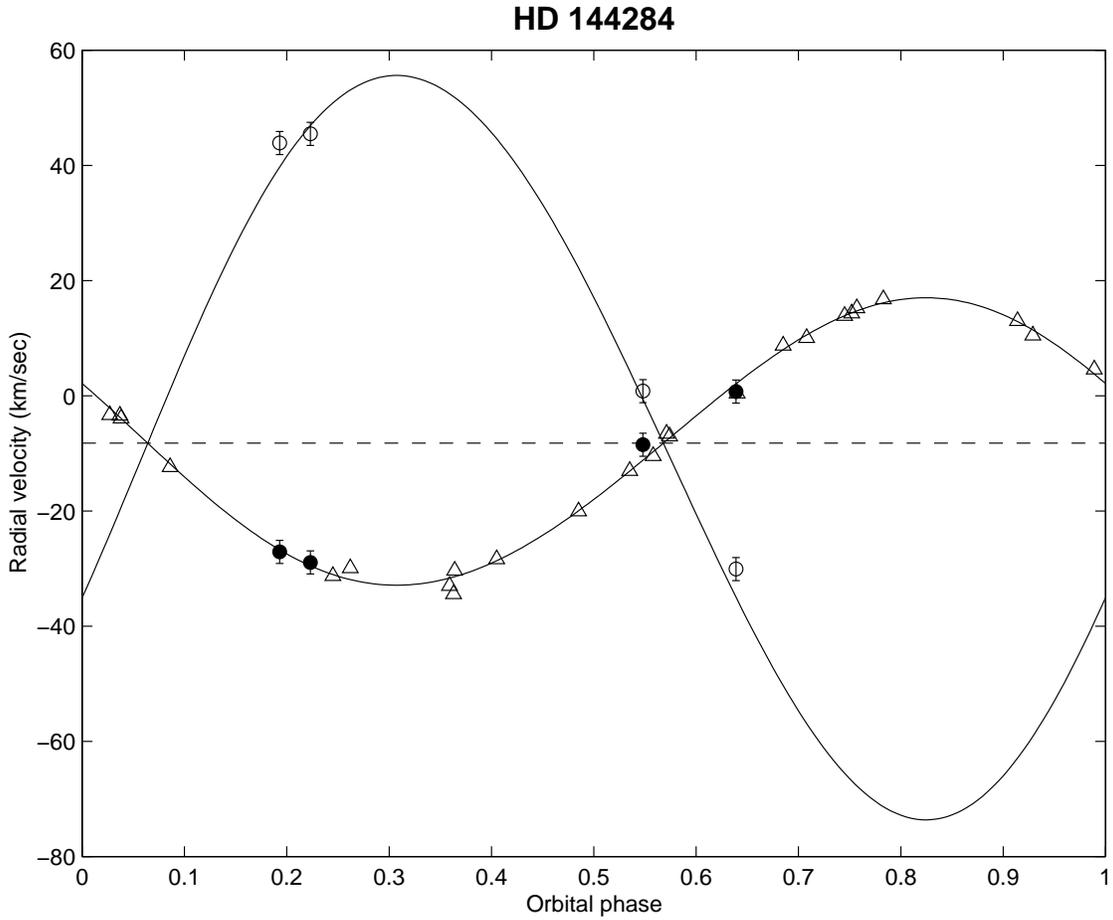}
\caption{Same as Figure 5 but for HD 144284.  Here the
previously measured velocities of the primary (open triangles) are
from \citet{duq91} and all the IR measurements were obtained using
CSHELL.}
 \end{figure}

\end{document}